\documentclass[pre,amsmath,amssymb,superscriptaddress,letterpaper]{revtex4}

\usepackage{graphicx}

\begin{document}

\title{Signal acquisition via polarization modulation in single photon sources}

\author{Mark D. McDonnell}
 \email{mark.mcdonnell@unisa.edu.au}
\affiliation{Institute for Telecommunications Research, University of South Australia, SA 5095, Australia}
\author{Adrian P. Flitney}
\affiliation{School of Physics, The University of Melbourne, VIC 3010, Australia}
\date{\today}

\begin{abstract}
A simple model system is introduced for demonstrating how a single photon source might be used to transduce classical analog information. The theoretical scheme results in measurements of analog source samples that are (i) quantized in the sense of analog-to-digital conversion and (ii) corrupted by random noise that is solely due to the quantum uncertainty in detecting the polarization state of each photon. This noise is unavoidable if more than one bit per sample is to be transmitted, and we show how it may be exploited in a manner inspired by suprathreshold stochastic resonance. The system is analyzed information theoretically, as it can be modeled as a noisy optical communication channel, although unlike classical Poisson channels, the detector's photon statistics are binomial. Previous results on binomial channels are adapted to demonstrate numerically that the classical information capacity, and thus the accuracy of the transduction, increases logarithmically with the square root of the number of photons, $N$. Although the capacity is shown to be reduced when an additional detector nonideality is present, the logarithmic increase with $N$ remains.
\end{abstract}

\maketitle

Extensive efforts to  establish practical quantum computers incorporating quantum communication and/or cryptography have led towards the development of reliable Single Photon Sources (SPS)~\cite{Keller.04,Strauf.07}. While much research in this area is targeted towards quantum information processing, and the concept of qubits~\cite{Nielson}, the impending availability of a reliable SPS may also have broader application.

Here we introduce and analyze a paradigm of quantum photonic information processing that falls beyond the scope of classical Poisson light~\cite{Mandel.76,Pierce.81,McEliece.81,Shamai.90}, sub-Poisson squeezed light~\cite{Saleh.87} and the usual focus of quantum information theory~\cite{Nielson}. We consider a model for the transduction of classical information via a quantum optical channel and a classical detector.

In our model, an information source directly modulates a SPS that emits photons with equal energy at a known constant rate. Unlike the Poisson or sub-Poisson statistics of a classical optical source, this results in {\em deterministic} photon release {\em counts}. Significant recent experimental progress~\cite{Strauf.07,Lukishova.07} indicates that an SPS may soon offer the possibility of modulation of the information source onto quantum properties of individual photons, namely the polarization angle. Thus the focus of this paper is on quantifying the potential benefits from exploiting such a controllable SPS polarization angle in sensing applications.

We model the information source to be sensed (the `signal') using the standard information theoretic approach, i.e.~we treat the signal as a discrete random variable. While our system results in a quantum information channel, it is capable of transmitting classical information, as considered, e.g., in~\cite{Hausladen.96,Schumacher.97}. This is similar to the scenario in~\cite{Peres.91,Hausladen.96} where two photons are considered in the context of the Holevo bound~\cite{Nielson}. However it differs in that we specify that an arbitrary (but constant) number of photons, $N$, are used for a single measurement of the signal, where each photon is produced independently and {\em unentangled} from a SPS. Thus the $N$ photons produced can be considered as a product input state~\cite{King.01}.

The second part of the model signal transduction system is a classical photon detector. We assume several standard practical constraints, i.e.~that the `receiver' consists of a classical horizontal polarizer and detector capable only of {\em counting} the number of photons received in a fixed duration, $t_s$, as in~\cite{Pierce.81,McEliece.81}. We further assume that the source sample period is also equal to $t_s$, so that the source and receiver are  synchronized in time. We quantify performance information theoretically in bits per source sample or bits per photon.

While practical photon counters are limited by inefficiency in the conversion of photons to electrons, and the photon count is a random variable for a given number that are transmitted~\cite{Romestain.04}, we consider only an ideal photon counter. We make this assumption because we aim to find the {\em upper limit} to the sensing accuracy of the scheme.  Other important losses in a practical system will lead to lesser performance; our results can be interpreted as a quantification of the `best case scenario,' which is the typical utility of information theory.

We first consider a relatively trivial scheme. Two ways of achieving one bit per source sample in time window $t_s$ include either transmitting or not transmitting a photon -- in which case the polarizer is unnecessary -- or to transmit either horizontally or vertically polarized photons. This may be viewed as a limiting case of classical on-off keyed (OOK) modulation. Clearly, this approach would be limited by the controllability of the SPS, and the readout/reset speeds of the detector. It may not be feasible to readout the photon count at the same rate as photons are produced, or it may only be feasible to produce $N>1$ equally polarized and unentangled photons during time $t_s$. If only horizontally or vertically polarized photons are used, this would be inefficient and redundant, and reduce the rate to $1/N$ bits per photon (although remaining equal to $t_s^{-1}$ bits per second.)

We now consider whether the number of bits per photon can be improved upon if the polarization angle during each sample period, $t_s$, can be set to any desired angle between vertical and horizontal. In this case, detection or non-detection of transmitted photons is no longer deterministic, and has a probability that depends on the original polarization angle.  However the actual photon count can be any of the integers in the range $0$,..,$N$, a fact that can be exploited to obtain $O\left(\frac{\log(N)}{N}\right)$ rather than $1/N$ bits per photon.

We assume that in a duration $t_s$ the SPS produces $N$ unentangled photons all perfectly polarized to some real-valued angle to horizontal, $\theta$. Let the state of a single photon polarized to angle $\theta$ be the qubit
\begin{equation}\label{qubit}
    |\psi\rangle = \cos{(\theta)}|H\rangle + \sin{(\theta)}|V\rangle,
\end{equation}
where $|H\rangle$ represents horizontally polarized and $|V\rangle$
represents vertically polarized.

Let the polarization angle of the detector relative to horizontal be the independent continuous random variable $\Phi$, with support $S_{\Phi}=[0,a]$, $a\in[0,\frac{\pi}{2}]$. When each photon passes through the
polarizer, the probability that it is
detected, conditioned on $\theta$, and $\Phi=\phi$ is
\begin{equation}\label{P_d}
    P_d(\theta|\phi) = \cos^2{(\theta-\phi)},\quad \theta\in\left[0,\frac{\pi}{2}\right].
\end{equation}

Let the density of $\Phi$ be $f_{\Phi}(\phi)$. We define $F_c(a)=\int_0^af_{\Phi}(\phi)\cos{(2\phi)}d\phi$ and $F_s(a)=\int_0^af_{\Phi}(\phi)\sin{(2\phi)}d\phi$. From Eq.~(\ref{P_d}), the probability of detecting any individual photon with polarization angle $\theta$ is given by
\begin{align}\label{P_dd}
    \bar{P_d}(\theta,a) &= \int_{S_{\Phi}}f_{\Phi}(\phi)P_d(\theta|\phi)d\phi\notag\\
    &=   0.5\left(1+\cos{(2\theta)}F_c(a)+\sin{(2\theta)}F_s(a)\right).
\end{align}

Let $y$ be the random variable representing the count of photons that
propagate through the horizontal polarizer after $N$ are transmitted with
polarization angle $\theta$, and let the distribution of source samples have a cumulative distribution function (CDF), $F_\theta(\theta)$. For $N>1$, the channel becomes a {\em binomial channel}~\cite{Komninakis.01}, since as the photons are unentangled, the probability that $n$ are counted at the detector is given by the binomial distribution,
\begin{equation}\label{P_y}
    P_{y|\theta}(y=n|\theta,a) = \left(^N_n\right)\bar{P_d}(\theta,a)^n(1-\bar{P_d}(\theta,a))^{(N-n)},
\end{equation}
where $n=0,..,N$. The mean number of photons counted for angle $\theta$ will be $\bar{N}=N\bar{P_d}(\theta,a)$.

The information theoretic limits of the performance of this channel can be found by maximizing the {\em mutual information}, $I(\theta;y)$, between the random variables $\theta$ and $y$~\cite{Gallager}. The result is the {\em channel capacity},
\begin{equation}\label{ChannelCapacityDef}
    C := \max_{\{F_\theta(\theta)\}} I(\theta;y).
\end{equation}
The units of $C$ and $I(\theta;y)$ are {\em bits per source sample}, where here each sample results in the emission of $N$ photons all polarized to the same angle $\theta$. Converting to bits per photon means dividing $C$ or $I$ by $N$, and to bits per second means dividing by $t_s$. Note the channel of Eq.~(\ref{P_y}) with {\em fixed} $N$ is a different kind of photonic `binomial channel' to that considered in~\cite{Mandel.76}, where the {\em number} of photons produced at a source is modeled as a controlled discrete random variable, $N$, with different values representing different source symbols, while photon detection probabilities are constant. In contrast, here $N$ is a constant and the photon detection probability is modulated by the angle $\theta$.

Many channel capacity problems include constraints on the source distribution~\cite{Gallager}. For example,~\cite{Shamai.90} considers maximum amplitude and average power constraints for standard Poisson light. Here we have both constraints implicitly, via the maximum number of photons per sample, $N$. Furthermore, we now show that the mutual information between $\theta$ and $y$ is equal to that between the variable $t:=\bar{P_d}(\theta,a)$ and $y$, and that allowing for uncertainty in the receiver (i.e.~$a>0$) is equivalent to placing {\em additional} amplitude constraints on $t$.

It is straightforward to show that $t=\bar{P_d}(\theta,a)$ has a single maximum for $\theta\in[0,\pi/2]$,
\begin{equation}\label{atan}
    \theta^o(a) := 0.5\arctan{\left(\frac{F_s(a)}{F_c(a)}\right)}.
\end{equation}
This attains maximum probability of detection
\begin{equation}\label{tmax}
    t_{\max}(a) := \bar{P_d}(\theta^o,a) = 0.5(1+\sqrt{F_c(a)^2+F_s(a)^2})\le 1.
\end{equation}
We also have at the extremes of horizontally or vertically polarized photons $\bar{P_d}(0,a) = 0.5(1+F_c(a))$ and $\bar{P_d}(\pi/2,a) = 0.5(1-F_c(a))$, and the minimum probability of detection is $t_{\min}(a) := \min\{\bar{P_d}(0,a),\bar{P_d}(\pi/2,a)\}\ge 0$.

The convexity of $t$ with respect to $\theta$ also implies that for $a>0$ some polarization angles induce the same conditional output distribution---that is, they produce ambiguous outcomes at the receiver. However, it also means that on the interval $t\in[t_{\min},t_{\max}]$ there exists a one to one continuous and invertible deterministic mapping $t\leftrightarrow\theta$. That is, $t$ is a monotonic function of $\theta$. Thus, we must have $I(\theta;y)=I(t;y)$, and the channel is equivalent to one with transition probabilities
\begin{equation}\label{P_y_t}
    P_{y|t}(y=n|t) = \left(^N_n\right)t^n(1-t)^{(N-n)},~t\in[t_{\min}(a),t_{\max}(a)].
\end{equation}
If polarization angles on the full interval $[0,\pi/2]$ are used, then any pair of angles that produce the same $t$ are interchangeable, and can be treated as the same input sample.

The capacity of the binomial channel of Eq.~(\ref{P_y_t}) (for $a=0$, i.e. no constraints on $t$) has been considered in several contexts,~\cite{Xie.97,Komninakis.01,McDonnell.07,McDonnell.09PRE}, and it is known that in the large $N$ limit
\begin{equation}\label{I_F}
    \lim_{N\rightarrow\infty}C = 0.5\log_2{\left(\frac{N\pi}{2e}\right)} := C_{\infty}.
\end{equation}
This is a lower bound to the capacity of the binomial channel of Eq.~(\ref{P_y_t}) when $N$ is finite~\cite{Komninakis.01}. Straightforward adaptation of analysis in~\cite{McDonnell.07} shows that this asymptotic capacity, $C_{\infty}$, is achieved when $\theta$ is a continuous random variable uniformly distributed between $0$ and $\pi/2$.

We now address the problem of  finding channel capacity for finite $N$ and $a\ge 0$. Since we have a discrete memoryless channel, with an output consisting of $N+1$ states, capacity is achieved when only a finite number of angles, $M\le N+1$, are used~\cite[Corollary~3,~pp.~96]{Gallager}. This means we can label the capacity achieving input distribution for the channel of Eq.~(\ref{P_y_t}) as a probability mass function, with real valued support given by $\{t_1,..,t_m,..,t_M\}$, and mass values $\{P_\theta(1),..,P_\theta(m),..,P_\theta(M)\}$. Since the transition probabilities are dependent on the support points, the capacity problem can be written as
\begin{equation}\label{ChannelCapacityDef1}
    C = \max_{\{M,P_\theta(1),..,P_\theta(m),t_1,..,t_M\}} I(t;y),
\end{equation}
where $M>1$, $\sum_{m=1}^MP_\theta(m) = 1$ and $t_m\in [0,1]$. Although mutual information is convex in the probabilities, $P_\theta(m)$, it is not convex in $t$ and solving using standard gradient descent methods does not guarantee a global optima.  However, several numerical approaches to solving such capacity problems exist. These all rely on the following property. Define $i(t_m):=\sum_{n=0}^NP(n|t=t_m)\log_2{\left(\frac{P(n|t=t_m)}{P_y(n)}\right)}$, where $P_y(n) = \sum_{m=1}^MP_\theta(m)P_{y|t=t_m}(n|t=t_m)$. By~\cite[Theorem 4.5.1]{Gallager}, a necessary and sufficient condition for achieving channel capacity for any given set of fixed $t_m$ of size $M$ is that
\begin{equation}\label{Nec}
    i(t_m) \left\{\begin{array}{cc}
                              = C & \forall~t_m~\mbox{s.t.}~P_\theta(m)>0 \\
                             \le C & \forall~t_m~\mbox{s.t.}~P_\theta(m)=0.
                          \end{array}\right.
\end{equation}
Consequently capacity is achieved if and only if all maxima of the function $i(t)$ are achieved by a support point $t=t_m$ and all support points achieve a maxima. This fact allows convex optimization techniques to be used, thus guaranteeing global convergence~\cite{Komninakis.01,McDonnell.09PRE}.  We used \texttt{CVX}, a package for specifying and solving convex programs \cite{Grant.09}.

\begin{figure}[ht]
\centerline{\includegraphics[scale=0.8]{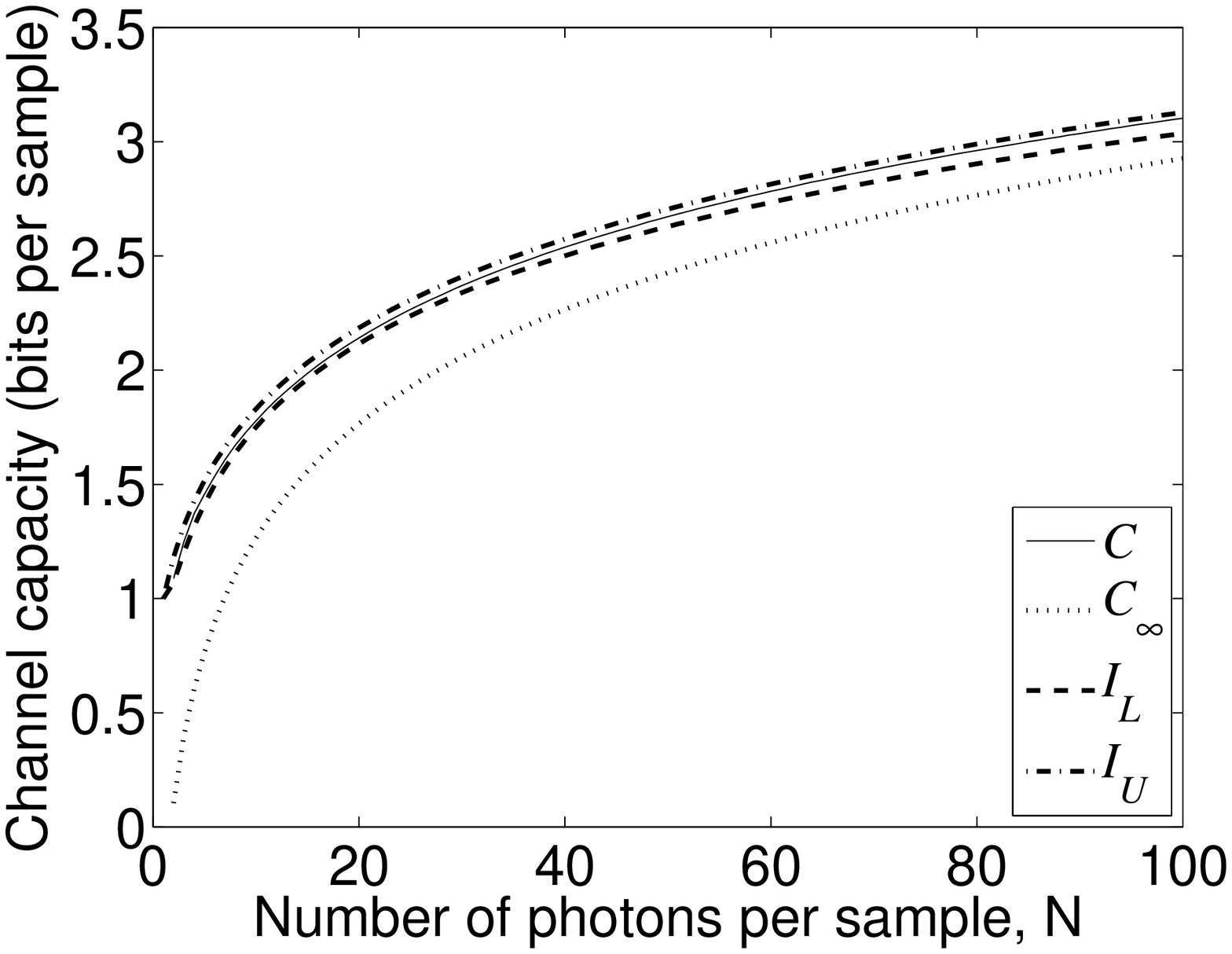}}
\caption{Channel capacity, $C$, for a perfect detector ($a=0)$, as a function of increasing number of photons per source sample, $N$ ($N$ is defined only for integer values, but a solid line is used as an aid to visibility). Also shown is $C_\infty$, i.e.~the lower bound to the channel capacity given by Eq.~(\ref{I_F}), which asymptotically approaches capacity as $N\rightarrow\infty$, a better lower bound to the channel capacity, $I_L$ (described in the text), and an upper bound, $I_U$.}\label{f:1}
\end{figure}

\begin{figure}[ht]
\centerline{\includegraphics[scale=0.8]{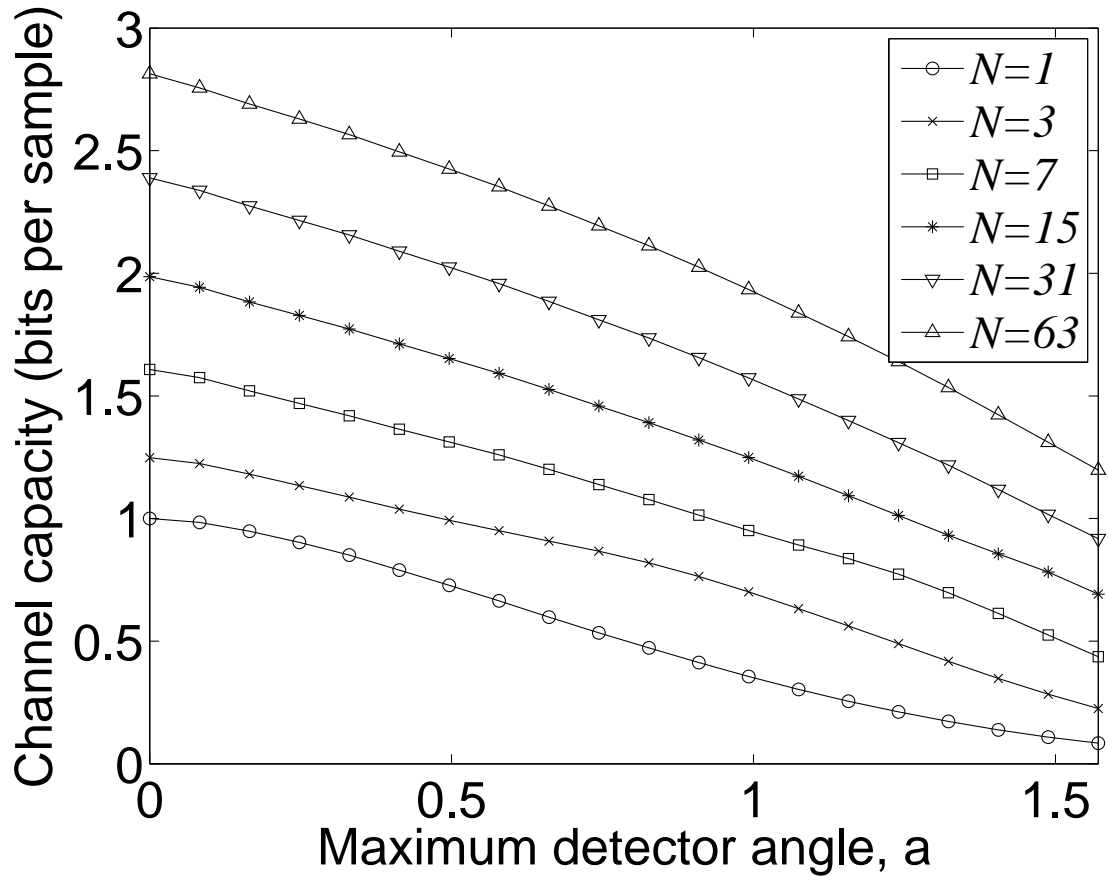}}
\caption{Channel capacity for various $N$ as a function of the maximum angle, $a$, of a randomly varying (uniformly distributed $\psi$) angle in the detector's polarizer (i.e.~increasing detector noise).}\label{f:2}
\end{figure}

The result of such a numerical optimization for this channel is shown with a solid line in Fig.~\ref{f:1}, for the case where the detector's angle is deterministic ($a=0$). The results are in agreement with~\cite{Komninakis.01,McDonnell.09PRE}. The capacity as a function of $a$, for various values of $N$ is shown in Fig.~\ref{f:2} for uniformly distributed detector angles. In this case $t_{\max}(a) = 0.5\left(1+\sin{(a)}/a\right)$ and $t_{\min}(a)=0.5\left(1-\sin{(2a)}/2a\right)$. While clearly decreasing as $a$ increases (which can be thought of as increasing detector noise) the capacity still increases with increasing $N$, and is larger than one bit when $N=63$ even at $a=0.5\pi$.

Fig.~\ref{f:1} also shows two known lower bounds to the capacity. The first, $C_\infty$ (given by Eq.~(\ref{I_F})) significantly underestimates the true capacity. The second, given in closed form in Eq.~(13) in~\cite{McDonnell.09PRE} and denoted here as $I_L$, is much tighter. In the photonic binomial channel context, it is equivalent to the mutual information that results by ensuring the source polarization angle has a mixture distribution that is absolutely continuous and uniformly distributed on $\left(0,\frac{\pi}{2}\right)$, and has discrete mass points at angles $0$ and $\frac{\pi}{2}$, each with some probability $Q<0.5$. This means that while the capacity is achieved by a purely discrete distribution, nearly the same performance can be achieved using analog modulation over all polarization angles.

Further verification can be obtained as follows. If an arbitrary output distribution $P_y(n)$ is chosen, the global maximum of $i(t)$ is an upper bound to capacity~\cite{Lapidoth.03}. Finding a close upper bound requires a judicial choice of $P_y(n)$. As an example, let
\begin{equation}\label{PyU}
    P_y^U(n) = \left\{\begin{array}{cc}
                     \frac{R}{1+2(R-Q)} & n=0,N \\
                     \frac{\left(^N_n\right)B(n+0.5,N-n+0.5)}{\pi(1+2(R-Q))}  & n=1,..,N-1.
                   \end{array}
    \right.
\end{equation}
where $B(\cdot,\cdot)$ is a beta function,
$Q = \frac{\Gamma(N+0.5)}{\sqrt{\pi}\Gamma(N+1)}$, and $R = \sqrt{\frac{2e}{N\pi}}$. Following~\cite{Lapidoth.03} we write
\begin{equation}\label{I_U}
I(x;y) \le I_U := \max_t\sum_{n=0}^NP(n|t)\log_2{\left(\frac{P(n|t)}{P_y^U(n)}\right)},
\end{equation}
where $P(n|t)$ is given by Eq.~(\ref{P_y_t}). The upper bound $I_U$ can be easily calculated numerically, and is shown in Fig.~\ref{f:1}, and as can be seen, it verifies the numerical calculation of capacity.

Our results mean that the achievable mutual information increases with $N$ as $O(0.5\log_2{(N)})$---c.f.~~Eq.~(\ref{I_F}). However since the number of bits per photon decreases to zero as $N$ increases, exploiting single photon polarization is of potential benefit in terms of bits per second only when the source sample does not change before more than one identically polarized photon is emitted. This fact may also be exploited in a hypothetical multi-photon source that releases $N$ identically polarized photons simultaneously. In this case, capacity is increased in an analogous way to utilizing multilevel pulse amplitude modulation (PAM) in conventional communication systems.  The difference is that the choice of the number of PAM amplitudes depends on channel noise, whereas for the SPS it depends on $N$, and indirectly on the quantum uncertainty in polarized photon detection.

Indeed, the model described here can be thought of as benefitting from randomness, in a similar sense to {\em suprathreshold stochastic resonance} (SSR)~\cite{Stocks.Mar2000,McDonnell.07}. The  model of~\cite{Stocks.Mar2000,McDonnell.07} has the same channel as Eq.~(\ref{P_y}), except $\bar{P_d}(\theta,a)$ is replaced by the CDF of {\em additive noise}. Mathematical equivalence (for $a=0$) follows, since photon detection is a random process equivalent to the corruption of $\theta$ by additive random noise, $\eta\in\left[-\frac{\pi}{4},\frac{\pi}{4}\right]$, with probability density function $f_\eta(\eta) =\cos{(2\eta)}$, followed by binary threshold detection such that the output is $|H\rangle$ when $\theta+\eta<\frac{\pi}{4}$. However, unlike~\cite{Stocks.Mar2000,McDonnell.07}, here SSR cannot be observed since the randomness is not due to noise, and the variance of $\eta$ cannot be changed (SSR might instead be observed if {\em sub-optimal} input signals were used). Nevertheless, it is only the uncertainty in the polarization state of each photon that allows an $N$ state quantization (digitization) of each sample, rather than a binary outcome. While a noisy signal results, like SSR this is offset by gaining more than one bit per source sample.

A significant outcome of our analysis is that we have shown that continuously distributed polarization angles can achieve performance close to capacity. Thus, the system may be highly suitable for the transduction of analog information sources. Indeed, if each photon is thought of as a ``noisy sensor,'' our hypothetical system bears similarities to several macroscopic frameworks studied in the signal processing literature---i.e.~the ``refining sensor network'' of~\cite{Gastpar.06}, or the ``stochastic pooling network'' of \cite{McDonnell.09PRE}.

\begin{acknowledgments}
Mark D. McDonnell is funded by the Australian Research Council, Grant No.~DP0770747. We thank Alex J. Grant, Terrence H. Chan and Nick Letzepis for valuable discussions.
\end{acknowledgments}

\clearpage

\clearpage

\end{document}